\documentclass{article}

\usepackage{PRIMEarxiv}
\usepackage{amsmath,amssymb,amsfonts}
\usepackage{algorithmic}
\usepackage[utf8]{inputenc} 
\usepackage[T1]{fontenc}    
\usepackage{hyperref}       
\usepackage{url}            
\usepackage{booktabs}       
\usepackage{amsfonts}       
\usepackage{nicefrac}       
\usepackage{microtype}      
\usepackage{lipsum}
\usepackage{fancyhdr}       
\usepackage{graphicx}       
\graphicspath{{media/}}     
\usepackage{algorithm} 

\title{Zero-Trust Runtime Verification for Agentic Payment Protocols: Mitigating Replay and Context-Binding Failures in AP2
\thanks{\textit{\underline{Citation}}: 
\textbf{Authors. Title. Pages.... DOI:000000/11111.}} 
}

\author{
 Qianlong Lan, Anuj Kaul, Shaun Jones, Stephanie Westrum
 \\
  eBay Inc \\
  \texttt{\{
 qialan, anukaul, shaujones, stwestrum\}@ebay.com} \\
}

\begin{document}
\maketitle

\begin{abstract}
The deployment of autonomous AI agents capable of executing commercial transactions has motivated the adoption of mandate-based payment authorization protocols, including the Universal Commerce Protocol (UCP) and the Agent Payments Protocol (AP2). These protocols replace interactive, session-based authorization with cryptographically issued mandates, enabling asynchronous and autonomous execution. While AP2 provides specification-level guarantees through signature verification, explicit binding, and expiration semantics, real-world agentic execution introduces runtime behaviors such as retries, concurrency, and orchestration that challenge implicit assumptions about mandate usage.

In this work, we present a security analysis of the AP2 mandate lifecycle and identify enforcement gaps that arise during runtime in agent-based payment systems. We propose a zero-trust runtime verification framework that enforces explicit context binding and consume-once mandate semantics using dynamically generated, time-bound nonces, ensuring that authorization decisions are evaluated at execution time rather than assumed from static issuance properties.

Through simulation-based evaluation under high concurrency, we show that context-aware binding and consume-once enforcement address distinct and complementary attack classes, and that both are required to prevent replay and context-redirect attacks. The proposed framework mitigates all evaluated attacks while maintaining stable verification latency of approximately 3.8~ms at throughput levels up to 10{,}000 transactions per second. We further demonstrate that the required runtime state is bounded by peak concurrency rather than cumulative transaction history, indicating that robust runtime security for agentic payment execution can be achieved with minimal and predictable overhead.
\end{abstract}


\section{Introduction}
Autonomous AI agents are increasingly entrusted with executing complex tasks on behalf of users, including financial transactions. Contemporary agent architectures frequently employ reasoning and action frameworks such as ReAct \cite{yao2022react} and tool use mechanisms that allow agents to invoke external services without direct human intervention \cite{schick2023toolformer}. As these agents operate with growing autonomy, traditional interactive payment models become inadequate, motivating the development of mandate-based payment authorization protocols, including the Universal Commerce Protocol (UCP) \cite{googledevs_ucp_2026} and the Agent Payments Protocol (AP2) \cite{googlecloud_ap2_2025}.

Unlike browser-centric payment standards, such as the World Wide Web Consortium (W3C) secure payment confirmation framework \cite{w3c_secure_payment_confirmation_2023}, which rely on continuous user presence and session-based authorization, AP2 is designed for headless agent execution. In these environments, payment authorization must function without persistent user interaction or synchronized approval flows. Consequently, agentic payment execution differs fundamentally from conventional e-commerce systems: it is stateless \cite{mahalle2022identity}, asynchronous, and highly concurrent, reflecting the operational characteristics of distributed agent runtimes.

To support this execution model, AP2 replaces session-based authorization with cryptographically issued mandates that delegate execution rights bound to specific transaction parameters. At the protocol level, these mandates provide well-defined security guarantees, including integrity, authenticity, and expiration semantics. However, these guarantees are derived under assumptions that abstract away the runtime behavior of autonomous agents. As a result, specification-level correctness does not necessarily imply secure execution in realistic deployment settings.

In practice, agentic systems routinely exhibit behaviors such as automatic retries, parallel task execution, inter-agent delegation, and observability-driven replay during debugging. These execution patterns introduce temporal and contextual complexity that was largely absent from earlier authorization models. Recent security analyses of Large Language Model (LLM) systems, including the Open Web Application Security Project (OWASP) Top 10 for LLMs \cite{owasp2023owasp} and studies on prompt injection and context manipulation \cite{lan2025prompt}, highlight how such dynamics can give rise to new attack surfaces. Within agentic payment systems, these conditions may enable payment mandates to be replayed, misapplied across execution contexts, or inadvertently exposed during runtime \cite{googledevs_ucp_2026}.

This paper examines the security implications of these runtime behaviors for the AP2 mandate lifecycle. We identify enforcement gaps that emerge when cryptographic mandates are consumed in highly concurrent and asynchronous agent environments and argue that additional runtime verification is required to uphold intended security properties. To address these challenges, we propose a zero-trust runtime verification framework that enforces explicit context binding and consume-once semantics at execution time. Unlike prior work that treats mandates as static authorization artifacts, this work systematically analyzes and enforces mandate execution semantics under the specific constraints of agentic concurrency and orchestration.

Our contributions are as follows:
\begin{itemize}
\item We present a systematic threat model for AP2 mandate execution in autonomous agent runtimes.
\item We identify concrete runtime enforcement gaps between AP2’s protocol specification and real-world deployments.
\item We introduce a zero-trust runtime verification framework that enforces context-aware binding and consume-once mandate semantics.
\item We evaluate the proposed framework under high-concurrency conditions, demonstrating effective mitigation of replay and context-binding attacks with minimal performance overhead.
\end{itemize}

\section{Background and Related Work}
\label{sec:headings}

\subsection{AP2 and Mandate-Based Authorization}
The Agent Payments Protocol (AP2) defines a mandate-based authorization model in which cryptographically issued tokens delegate limited execution rights to autonomous agents \cite{googlecloud_ap2_2025}. These mandates such as checkout mandates and payment mandates authorize specific actions within narrowly defined scopes, allowing agents to initiate and complete transactions without continuous user involvement.

From a structural perspective, AP2 mandates resemble JSON Web Tokens (JWTs) \cite{jones2015json} commonly used in OAuth 2.0 authorization flows \cite{jones2012oauth}. Each mandate encapsulates authorization metadata and is protected by digital signatures, key binding, and explicit expiration timestamps. Together, these mechanisms provide specification-level guarantees of integrity, authenticity, and temporal validity. Unlike interactive authorization schemes, however, mandates are intended to be consumed autonomously by agents operating outside the context of persistent user sessions.

\subsection{Agentic Execution Model}
Agentic payment execution departs significantly from traditional client–server authorization models. Autonomous agents are typically stateless, operate asynchronously, and execute tasks concurrently across distributed environments \cite{googledevs_ucp_2026}. Rather than following a single, linear control flow, agentic systems frequently rely on orchestration mechanisms that coordinate multiple subtasks and agents to achieve a higher-level objective.

These execution characteristics introduce behaviors such as automatic retries, parallel invocation of tools, and delegation across agents, particularly in the presence of partial failures or uncertain outcomes. While such behaviors are essential for robustness and scalability, they complicate assumptions commonly associated with authorization tokens most notably, that authorization is exercised exactly once and within a single, well-defined execution context. As a result, mandate consumption semantics that are sufficient in synchronous, session-oriented systems may not hold in agentic runtimes.

\subsection{Related Work}
The challenges posed by agentic execution align with broader trends in security architecture that question implicit trust derived from session state. Zero-trust security models \cite{nist_800_207} advocate continuous verification of authorization decisions rather than reliance on perimeter-based or session-bound trust assumptions \cite{nist_800_204}. Similarly, capability-based security systems emphasize fine-grained, explicitly scoped authorization tokens that can be independently validated at the point of use \cite{jones2015json,anderson2010security}.

Prior research has explored these principles in distributed systems, web security, and access control frameworks \cite{fett2016comprehensive}. In parallel, formal methods and runtime verification techniques have been developed to enforce safety and correctness properties during system execution \cite{sanchez_runtime_verification_survey_2019}. However, existing work has not examined how mandate-based authorization should be enforced in autonomous agent payment systems, where execution is asynchronous, highly concurrent, and driven by learned policies rather than fixed control logic. This gap motivates the present study. To the best of our knowledge, no prior work has specifically studied how authorization tokens behave under the unique combination of agent-driven retries, orchestration, and observability leakage in real-world payment execution.

\section{System and Threat Model}
\label{sec:threat_model}
\subsection{System Model}
We consider an agentic commerce environment comprising three primary entities: the \textbf{User Agent} ($\mathcal{A}$), the \textbf{Merchant Endpoint} ($\mathcal{M}$), and the \textbf{Runtime Verifier} ($\mathcal{V}$).

\begin{itemize}
    \item \textbf{User Agent ($\mathcal{A}$):} An autonomous software entity authorized to execute transactions on behalf of a user. It possesses a cryptographic mandate $M$ signed by a user authority.
    \item \textbf{Merchant Endpoint ($\mathcal{M}$):} The target service provider that fulfills the transaction. $\mathcal{M}$ relies on $\mathcal{V}$ to validate incoming requests.
    \item \textbf{Runtime Verifier ($\mathcal{V}$):} A trusted middleware component acting as a microservices security gateway complying with  NIST SP 800-204 \cite{nist_800_204} responsible for enforcing mandate usage policies.
\end{itemize}

We assume cryptographic primitives (e.g., digital signatures, hash functions) are secure \cite{katz2007introduction}. The Verifier $\mathcal{V}$ is part of the Trusted Computing Base (TCB). Network-level attacks (e.g., DDoS) and side-channel leakage \cite{konigsmark2017high} are considered out of scope. Compromise of the verifier itself is considered equivalent to a compromise of the underlying payment infrastructure and is therefore out of scope.

\subsection{Threat Taxonomy}
We define the following specific threats targeting the execution phase of agentic payments:

\begin{itemize}
    \item \textbf{T1: Same-Context Replay (Temporal Violation).}
    An attacker (or a misconfigured agent scheduler) resubmits a valid mandate $M$ to $\mathcal{M}$ multiple times within its validity window $\Delta t$. This often occurs during aggressive retry loops or race conditions.
    
    \item \textbf{T2: Cross-Context Replay (Spatial Violation).}
    A valid mandate $M$ intended for a specific task or merchant context $C_A$ is maliciously redirected to a different context $C_B$. This constitutes a \textit{Context Binding Failure}, where the cryptographic signature remains valid, but the execution intent is violated.
    
    \item \textbf{T3: Context Leakage Induced Misuse.}
Sensitive semantic information (e.g., item details, budget constraints, or execution scope) embedded in a mandate may be exposed to unauthorized entities through excessive data inclusion, logging, or observability pipelines. While such leakage constitutes a confidentiality concern, we focus on the resulting execution-level misuse, where leaked mandate material is replayed or redirected to authorize unintended actions.

    \item \textbf{T4: Observability-Based Replay.}
    Valid mandates are harvested from unencrypted observability pipelines (e.g., distributed traces, debug logs) and replayed by an adversary with access to the monitoring infrastructure.
\end{itemize}

\subsection{Specification-Level Guarantees vs. Runtime Gaps}
The AP2 specification provides strong static guarantees for mandates, including integrity via cryptographic signatures and time-bounded validity via Time-To-Live (TTL)/expiry. However, the specification does not normatively define verifier-side, stateful execution semantics that hold under real-world agent retries and orchestration. In particular, AP2 does not require (i) consume-once enforcement (e.g., idempotency keys or nonce registries) to prevent multiple executions within the validity window, nor (ii) strict execution-time context binding that ensures mandate consumption is tied to the intended merchant/payee when requests traverse shared payment-processor endpoints. Notably, the specification highlights ``Mandate--Merchant Matching'' and temporal gaps as risk considerations, suggesting that these checks are left to deployment-specific enforcement rather than guaranteed by the protocol.
These gaps between protocol design and runtime reality motivate the need for the Zero-Trust Runtime Verifier proposed in Section \ref{sec:design_ztrv}.

\section{Design of the Zero-Trust Runtime Verifier}
\label{sec:design_ztrv}
\subsection{Design Goals}
The framework is designed to achieve the following goals:
\begin{itemize}
    \item Prevent replay under retries and concurrency.
    \item Enforce explicit binding between mandates and execution contexts.
    \item Fail closed on any verification error.
    \item Avoid reliance on agent state or session continuity.
\end{itemize}

\subsection{Architecture Overview}

\begin{figure}[htbp]
\centerline{\includegraphics[width=5in]{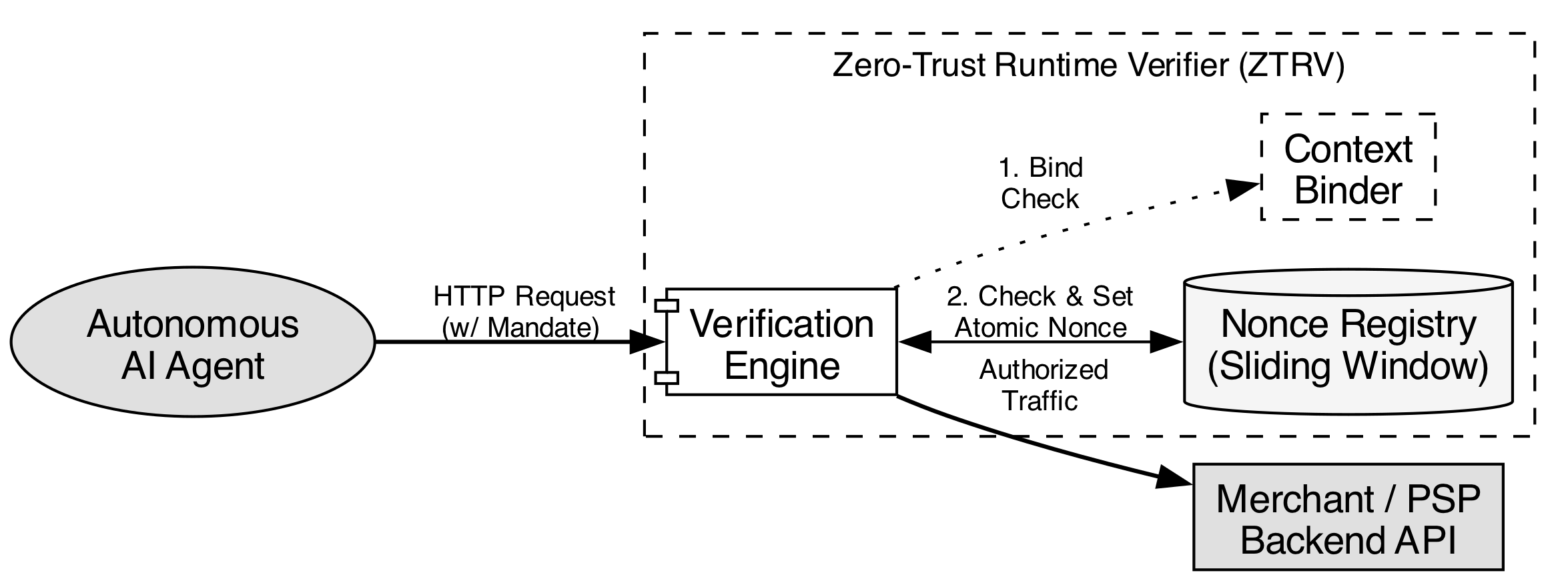}} 
\caption{Zero-Trust Runtime Verifier (ZTRV) architecture}
\label{fig0}
\end{figure}

We introduce the Zero-Trust Runtime Verifier (ZTRV), a mediation layer positioned between autonomous agents and merchant or payment service provider (PSP) backends, as illustrated in Figure~\ref{fig0}. The ZTRV enforces runtime security guarantees for mandate-based payment execution by requiring that all mandate-bearing requests undergo explicit verification prior to execution. Autonomous agents are not permitted to interact directly with merchant or PSP endpoints; instead, every request carrying an authorization mandate must be routed through the verifier.

At a high level, the ZTRV functions as a stateless verification gateway augmented with minimal, time-bounded state necessary to enforce consume-once semantics. Incoming HTTP requests from autonomous agents include a cryptographic mandate along with execution metadata. These requests are first received by the Verification Engine, which orchestrates all validation steps before forwarding authorized traffic to downstream services.

The verification pipeline comprises three stages. The Context Binder validates that the execution context of the incoming request is consistent with the constraints encoded in the mandate. This step enforces explicit binding between the mandate and relevant execution attributes such as transaction parameters, agent identity, and intended action scope thereby preventing reuse in unintended contexts.

Replay resistance is enforced through an atomic nonce check. Each mandate execution is associated with a dynamically generated, time-bound nonce that is registered upon first use. The Nonce Registry, implemented as a sliding-window data structure, ensures that each nonce is accepted at most once within its validity window, enabling consume-once semantics under high concurrency.

Requests that satisfy both checks are classified as authorized and forwarded to the merchant or PSP backend API for execution. Requests that fail verification are rejected before reaching the payment infrastructure, ensuring that enforcement occurs prior to any externally observable side effects.

\subsection{Context-Aware Binding}
Each mandate is bound to a runtime execution context through a cryptographic hash computed as:
\begin{equation}
H_{\text{ctx}} = \text{SHA256}(task.id \parallel agent.id \parallel merchant.id \parallel scope)
\end{equation}

 The exact set of context attributes is policy-configurable; the listed fields represent a minimal binding sufficient to prevent cross-task and cross-merchant reuse in our target scenarios. Upon receiving a request, the verifier reconstructs the context hash $H'{\text{ctx}}$ using attributes derived from the incoming execution environment, such as the authenticated agent identity, the target merchant endpoint, and the requested operation. The reconstructed hash is compared against the context hash $H{\text{signed}}$ embedded within the mandate’s cryptographic signature. Any mismatch ($H'{\text{ctx}} \neq H{\text{signed}}$) is treated as a context violation and results in immediate rejection of the request.

This binding mechanism ensures that mandates cannot be replayed across tasks, agents, or merchants, even if the underlying cryptographic token remains valid.

\subsection{Dynamic Consume-Once Nonce Registry}
To enforce single-use semantics and idempotent execution \cite{helland2012idempotence}, the ZTRV maintains a dynamic nonce registry. Each mandate includes a unique nonce that may be consumed exactly once during its validity period. The verifier performs an atomic check-and-set operation on the nonce registry for every incoming request to ensure correctness under concurrent execution \cite{kleppmann2019designing}.

The nonce registry is implemented using a sliding expiration window, allowing nonces to be garbage-collected automatically after a bounded time interval. This design achieves replay protection with minimal state overhead and is conceptually similar to distributed locking mechanisms such as Redlock \cite{redis_distributed_locks_2026}, while remaining optimized for high-throughput verification workloads.

\subsection{Runtime Verification Algorithm}
The Zero-Trust Runtime Verifier (ZTRV) enforces a strict, fail-closed verification pipeline for every incoming mandate-bearing request. Let $\mathcal{S}$ denote the ephemeral nonce registry and let $\Delta t$ represent the nonce validity window. The complete verification procedure, including context binding, nonce enforcement, and authorization forwarding, is formalized in Algorithm 1.

\begin{algorithm}
\caption{Zero-Trust Runtime Verification Logic}
\begin{algorithmic}[1]
\REQUIRE Incoming Request $R$, System Time $T_{now}$
\REQUIRE Window Size $\Delta t$, Nonce Registry $\mathcal{S}$
\ENSURE Transaction Decision $\{ACCEPT, REJECT\}$

\STATE \textbf{Step 1: Parse Mandate}
\STATE $M \leftarrow R.mandate$

\STATE \textbf{Step 2: Cryptographic \& Freshness Check}
\IF {$\neg \text{VerifySig}(M, R.signature)$}
    \RETURN $REJECT(\text{"Invalid Signature"})$
\ENDIF
\IF {$(T_{now} - M.timestamp) > \Delta t$}
    \RETURN $REJECT(\text{"Mandate Expired"})$
\ENDIF

\STATE \textbf{Step 3: Context Integrity (Anti-Leakage)}
\STATE $H'_{ctx} \leftarrow \text{SHA256}(R.merchant\_id \parallel R.path \parallel \dots)$
\IF {$H'_{ctx} \neq M.context\_hash$}
    \RETURN $REJECT(\text{"Context Mismatch"})$
\ENDIF

\STATE \textbf{Step 4: Atomic Consumption (Anti-Replay)}
\STATE $Key \leftarrow \text{"nonce:"} + M.nonce$
\STATE $LockAcquired \leftarrow \mathcal{S}.\text{SetNX}(Key, \text{TTL}=\Delta t)$
\IF {$\neg LockAcquired$}
    \RETURN $REJECT(\text{"Replay Detected"})$
\ENDIF

\STATE \textbf{Step 5: Final Authorization}
\RETURN $ACCEPT(\text{"Authorized"})$

\end{algorithmic}
\end{algorithm}

Requests failing any step are rejected without fallback, ensuring a secure-by-default execution posture.

\section{Experimental Evaluation}
This section evaluates the security effectiveness and performance characteristics of the proposed Zero-Trust Runtime Verifier (ZTRV). We compare ZTRV against a baseline verifier that reflects standard AP2-style mandate validation and assess both security outcomes under adversarial conditions and runtime overhead under increasing system load. 
\subsection{Experimental Setup}
We implemented the ZTRV as a standalone verification service in Python and evaluated it under simulated agent workloads representative of autonomous payment execution. The verifier processes HTTP requests carrying cryptographic mandates and associated execution metadata, performing runtime validation prior to forwarding authorized requests to a mock merchant backend.

As a point of comparison, we implemented a baseline verifier that performs standard AP2-style checks, including signature verification and expiration validation, but does not enforce consume-once semantics or context-aware binding. This baseline reflects the guarantees provided at the protocol specification level without additional runtime enforcement. It is important to note that this baseline reflects the AP2 protocol specification as written, strictly adhering to its cryptographic requirements without additional, non-standard execution-layer enforcement.

All experiments were conducted under controlled conditions with configurable concurrency and request rates to emulate realistic agent behavior, including retries and parallel execution.

\subsection{Attack Scenarios}
We evaluate the verifier under three attack scenarios derived from the threat model described in Section \ref{sec:threat_model}. These scenarios capture common classes of mandate misuse in agentic execution environments: same-context replay, cross-context replay and context-redirect attacks. Each scenario is executed under concurrent workloads to reflect realistic retry and orchestration behavior.

\subsection{Security Effectiveness}

\begin{table}
 \caption{Defense Effectiveness Against Context Leakage}
  \centering
  \begin{tabular}{lll}
    \toprule

     \textbf{Metric} & \textbf{Baseline (Standard AP2)} & \textbf{Proposed (ZTRV)}  \\
    \midrule
   \textit{Capability} & & \\
    Context Integrity Check & Absent & \textbf{Enforced} \\
    Consume-Once Semantics & Absent & \textbf{Enforced} \\
    \midrule
    
    \textit{Performance (N=5,000)} & & \\
   Interception Rate & $0.00\%$ & $\mathbf{100.00\%}$ \\
    False Positive Rate & N/A & $0.00\%$ \\
    \bottomrule
  \end{tabular}
  \label{tab1}
\end{table}

We focus on preventing execution-level misuse arising from leaked mandate material, rather than on confidentiality of mandate contents themselves.
Across all evaluated attack scenarios, the baseline verifier permits replay or misapplication of mandates when retries or concurrent execution occur. In contrast, the proposed ZTRV rejects all evaluated attacks by enforcing runtime context binding and consume-once semantics.

\begin{figure}[htbp]
\centerline{\includegraphics[width=5in]{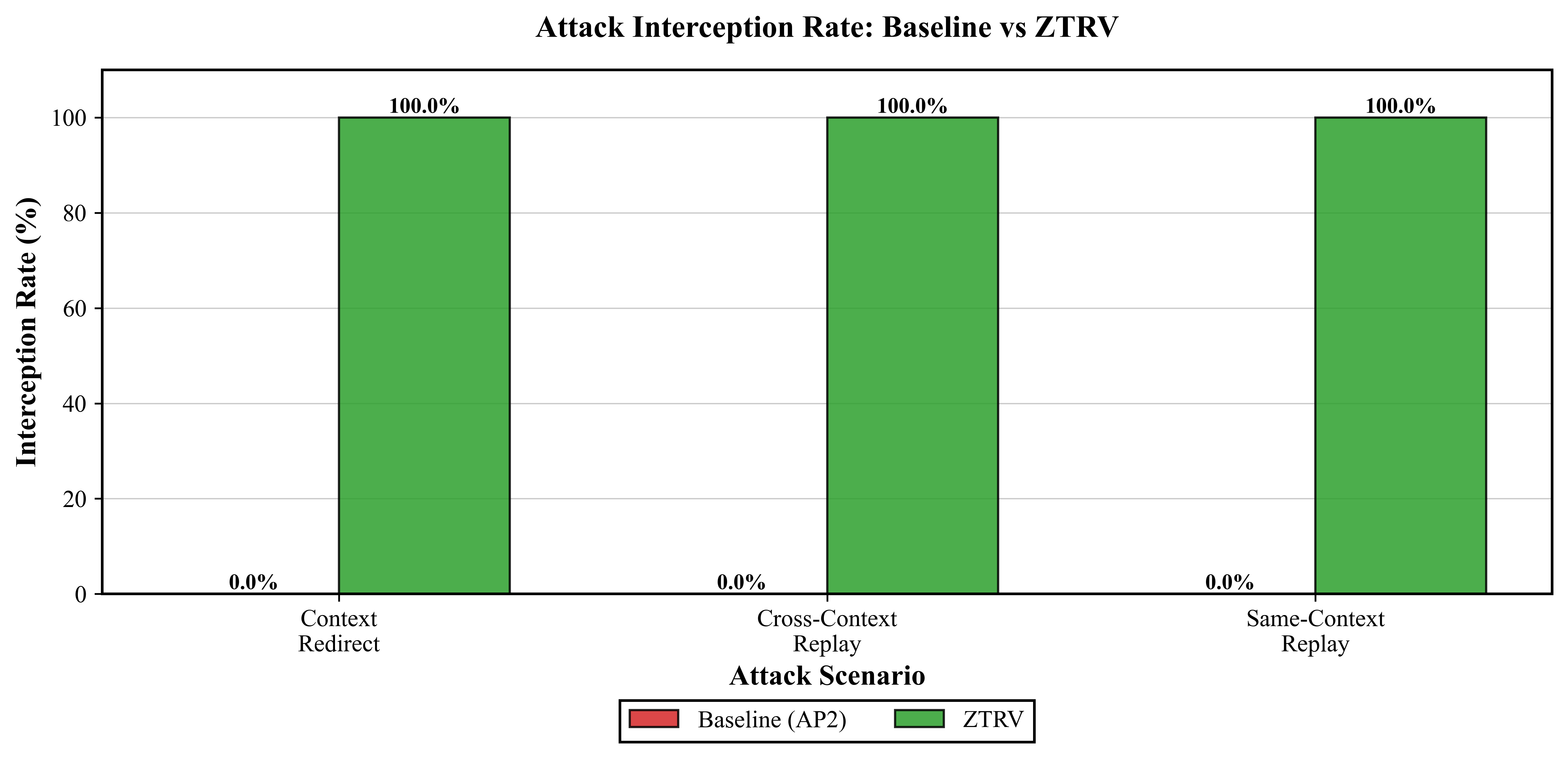}} 
\caption{Attack Interception Rate: Baseline vs. ZTRV.}
\label{fig1}
\end{figure}

Figure~\ref{fig1} summarizes the interception rate for each attack class. While the baseline verifier fails to intercept any of the tested attacks, the ZTRV achieves a 100\% interception rate across all scenarios.

The effectiveness of the ZTRV against context leakage attacks (Attack Vector B) is further detailed in Table~\ref{tab1}. The results show that the proposed framework enforces both context integrity and single-use semantics without introducing false positives, even under high concurrency.

\subsection{Performance Evaluation}
\begin{figure}[htbp]
\centerline{\includegraphics[width=5in]{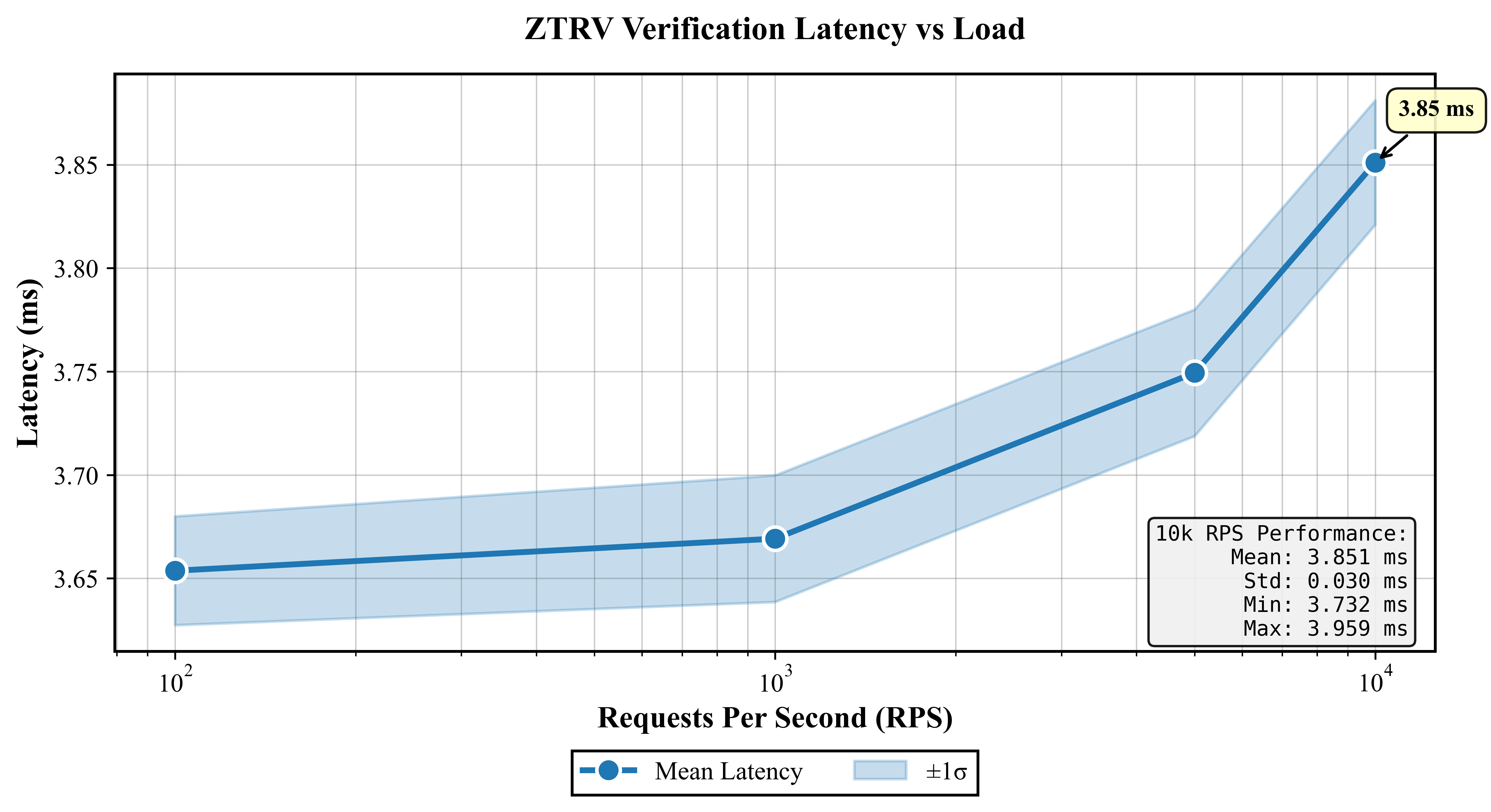}} 
\caption{Processing Latency under increasing system load.}
\label{fig2}
\end{figure}
We evaluate the runtime overhead introduced by the ZTRV under increasing system load. Figure~\ref{fig2} reports the average processing latency as a function of request throughput, measured in transactions per second.

Across all tested load levels, the verifier maintains stable performance. At a throughput of 10,000 transactions per second, the average verification latency is approximately 3.8 ms. The variance remains low, and no throughput bottlenecks are observed. These results indicate that the additional runtime checks introduced by ZTRV impose minimal overhead and scale effectively under high-concurrency conditions.

\subsection{Ablation Studies}
\label{sec:ablation}

To better understand the contribution of individual enforcement mechanisms and the scalability properties of the proposed verifier, we conduct a series of ablation studies. These experiments isolate the effects of context-aware binding and consume-once enforcement, examine the sensitivity of storage overhead to the nonce validity window, and decompose the runtime verification cost into its constituent components.
\subsubsection{Mechanism Isolation: Context Binding vs. Consume-Once}

\begin{figure}[htbp]
\centerline{\includegraphics[width=5in]{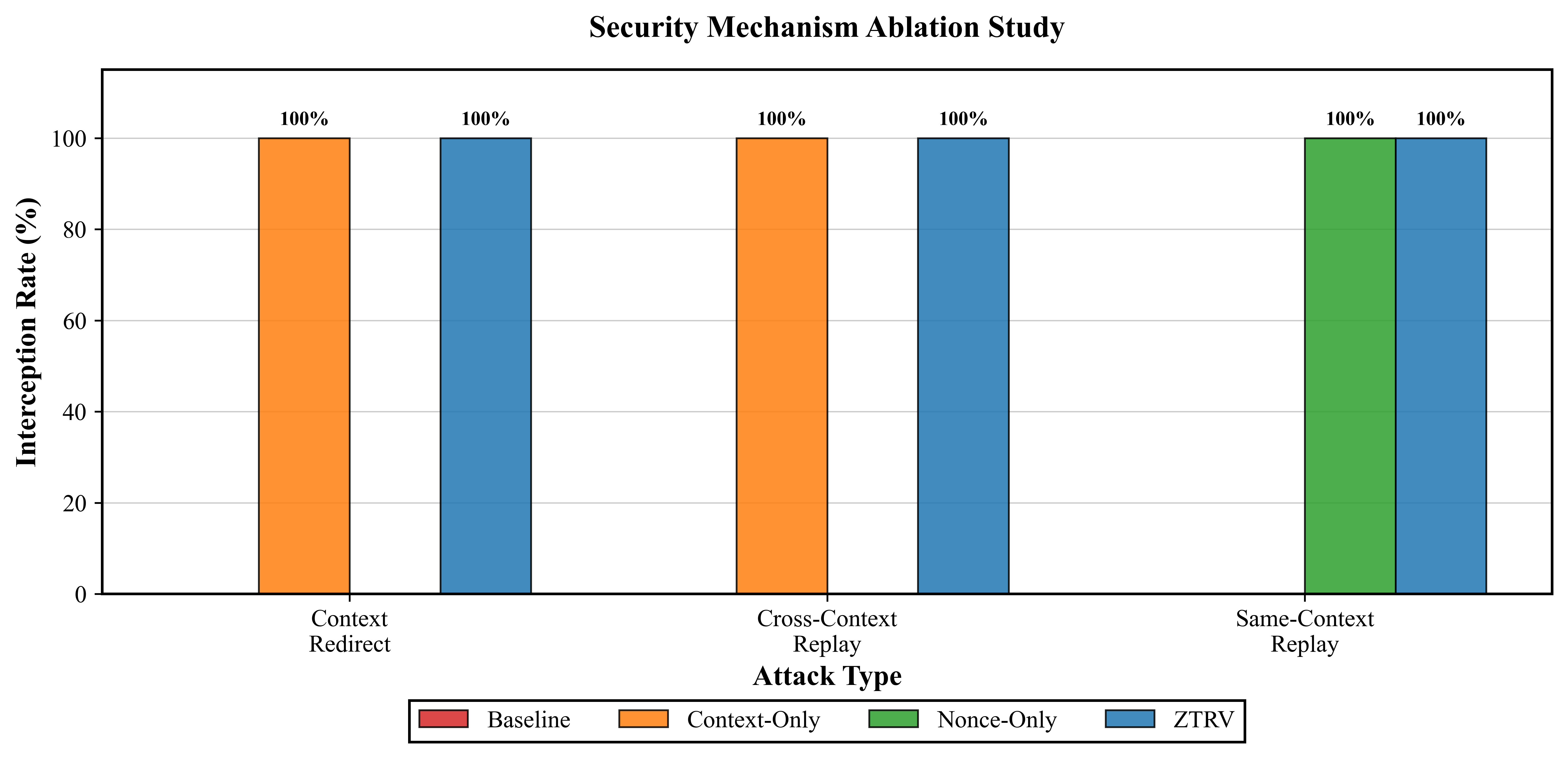}} 
\caption{Ablation study of attack interception rates under different verifier configurations. Context-aware binding and consume-once enforcement address complementary attack classes; only their combination (ZTRV) mitigates all evaluated attacks.}
\label{fig:ablation_attack_interception}
\end{figure}

We first evaluate the individual contribution of context-aware binding and consume-once enforcement by comparing four verifier configurations: (i) a baseline verifier performing only signature and expiration checks, (ii) a \textit{context-only} verifier that enforces context binding without nonce consumption, (iii) a \textit{nonce-only} verifier that enforces consume-once semantics without context binding, and (iv) the full ZTRV configuration combining both mechanisms.

Figure~\ref{fig:ablation_attack_interception} reports the interception rate across three attack scenarios. The results show that neither mechanism alone is sufficient to prevent all classes of mandate misuse. Consume-once enforcement alone successfully mitigates same-context replay attacks but fails to prevent cross-context replay and context-redirect attacks. Conversely, context binding alone prevents cross-context and redirect attacks but remains vulnerable to same-context replays. Only the full ZTRV configuration achieves complete interception across all evaluated scenarios.

These results demonstrate that context-aware binding and consume-once enforcement address distinct and complementary failure modes, and that both are required to ensure robust runtime security for agentic payment execution.

\begin{figure}[htbp]
\centerline{\includegraphics[width=5in]{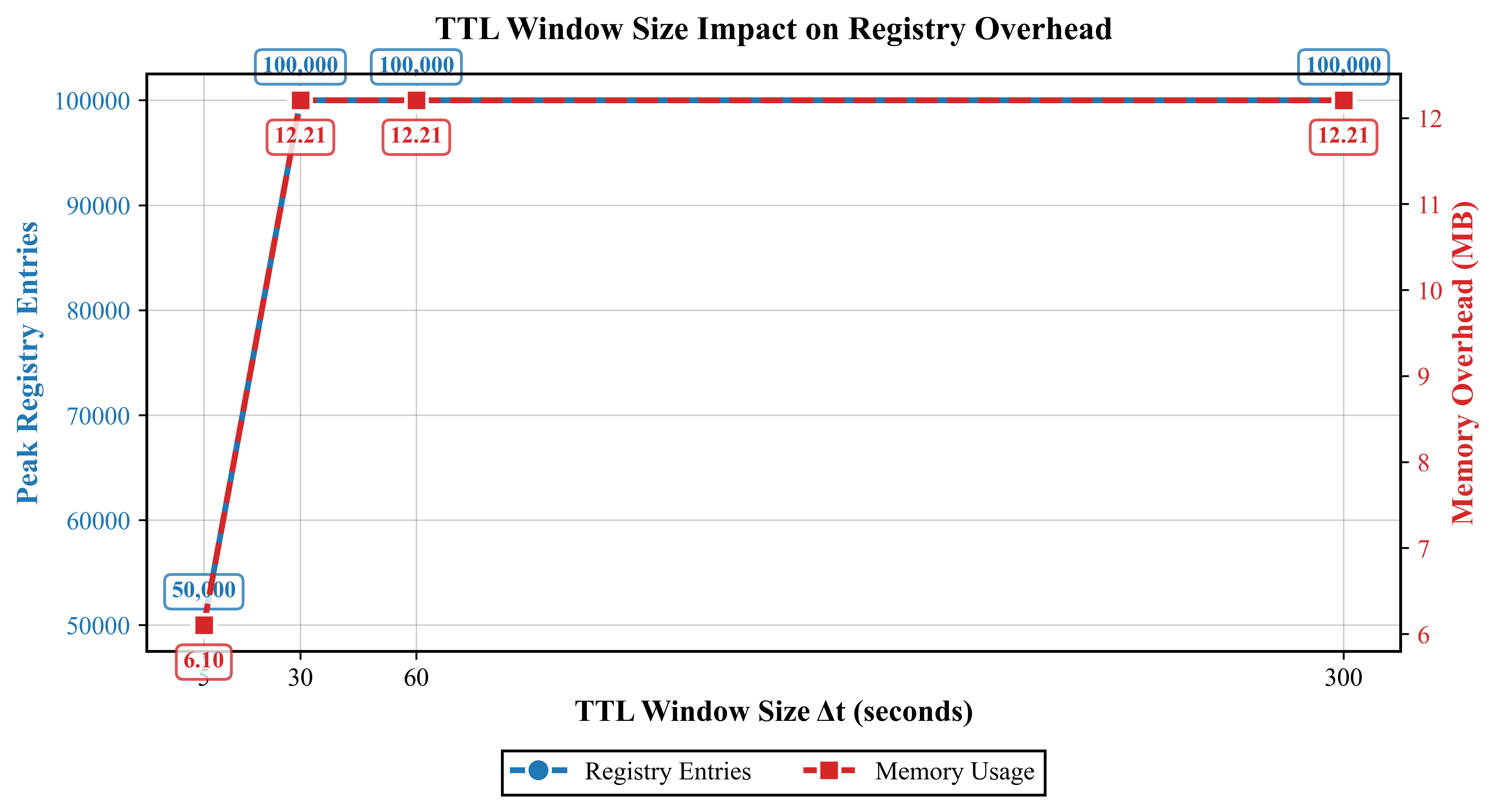}} 
\caption{Nonce registry size and estimated memory usage as a function of the nonce validity window ($\Delta t$) at 10{,}000 TPS. Storage scales with peak concurrency and plateaus once $\Delta t$ exceeds the experiment duration.}
\label{fig:ttl_sensitivity}
\end{figure}

\subsubsection{Nonce Window Sensitivity and Storage Overhead}

We next examine the impact of the nonce validity window ($\Delta t$) on the size of the nonce registry and associated storage overhead. We sweep $\Delta t \in \{5, 30, 60, 300\}$ seconds while fixing the request rate at 10{,}000 transactions per second and the experiment duration to 10 seconds.

Figure~\ref{fig:ttl_sensitivity} shows the peak number of active nonce entries and the corresponding estimated memory footprint. When the nonce window is shorter than the experiment duration, the registry size grows proportionally to $\Delta t$. Once $\Delta t$ exceeds the experiment duration, the registry size plateaus, reflecting that storage requirements are bounded by peak concurrency rather than cumulative transaction history.

Even at the largest evaluated window, the registry footprint remains modest (approximately 12~MB at peak), indicating that the proposed design scales efficiently under high-throughput workloads and is well suited for cloud-native deployments.

\subsubsection{Runtime Overhead Breakdown}

To clarify the source of runtime overhead introduced by ZTRV, we decompose the verification latency into three components: cryptographic signature verification, context hash validation, and nonce registry operations. Table~\ref{tab:overhead_breakdown} reports the average per-request cost of each component under increasing request rates.

Across all evaluated loads, signature verification dominates the total verification cost, while context binding incurs negligible overhead. Nonce registry operations contribute a moderate and stable cost that increases slightly under higher concurrency, consistent with contention effects in shared state management. The total verification latency remains stable at approximately 3.7~ms even at 10{,}000 requests per second, confirming that the additional runtime enforcement does not introduce performance bottlenecks.

This breakdown reinforces that ZTRV achieves strong security guarantees with minimal and predictable overhead.

\begin{table}[htbp]
 \caption{Runtime verification overhead breakdown under increasing system load. Signature verification dominates total cost, while context binding and nonce enforcement incur minimal additional overhead.}
  \centering
  \begin{tabular}{r|rrrr}
    \toprule
    RPS & Signature (ms) & Context (ms) & Registry (ms) & Total (ms) \\
    \midrule
    100 & 2.500 & 0.200 & 1.000 & 3.700 \\
    1000 & 2.502 & 0.199 & 1.001 & 3.703 \\
    5000 & 2.505 & 0.200 & 1.004 & 3.709 \\
    10000 & 2.509 & 0.200 & 1.010 & 3.720 \\
    \bottomrule
  \end{tabular}
  \label{tab:overhead_breakdown}
\end{table}

\section{Discussion and Limitations}

\subsection{Deterministic Security vs. Heuristics}
The experimental evaluation demonstrates a 100\% interception rate for all evaluated replay and context-redirect attacks (Figure~\ref{fig1}). It is important to distinguish this outcome from the performance of heuristic-based security mechanisms, such as AI-driven fraud detection systems, which rely on probabilistic inference and are therefore susceptible to false positives and false negatives \cite{proofpoint_threat_detection_2024}.

In contrast, the ZTRV framework enforces security properties through deterministic cryptographic logic. Authorization decisions are derived from explicit checks over mandate bindings and nonce state: if a nonce is already present in the registry, the corresponding request is rejected with certainty. This binary decision model explains the perfect detection rates observed in simulation, assuming the underlying storage layer provides strong consistency guarantees. As a result, ZTRV complements rather than replaces heuristic defenses by providing strict correctness guarantees at the execution layer.

\subsection{Storage Scalability \& TTL}
A common concern with runtime verification mechanisms is the potential growth of state under high transaction volume. In the case of ZTRV, this risk is mitigated through the use of time-bounded nonce storage. Each nonce is retained only for the duration of a predefined validity window ($\Delta t = 60,\mathrm{s}$), following established Time-To-Live (TTL) patterns in distributed caching systems \cite{redis_distributed_caching_2025}.

As a consequence, storage requirements scale with peak concurrency rather than cumulative transaction history. For example, at a sustained throughput of 10,000 transactions per second, the active nonce registry contains only the nonces issued within a single window, resulting in a memory footprint on the order of several megabytes. This bounded state model makes the approach well suited for cloud-native and horizontally scalable deployments.

\subsection{Implementation Challenges}
\subsubsection{Clock Synchronization}
A practical challenge in deploying time-bound mandate verification arises from clock drift, a well-known issue in distributed systems \cite{latha2010clock}. The verification algorithm compares timestamps embedded in mandates against the verifier’s local system time. If the verifier’s clock lags significantly behind that of the issuing agent, valid mandates may be incorrectly rejected.

To mitigate this risk, deployments must ensure reliable clock synchronization through mechanisms such as the Network Time Protocol (NTP). Alternatively, a bounded tolerance window (e.g., $\pm 5,\mathrm{s}$) can be incorporated into the verification logic to absorb minor synchronization discrepancies without weakening security guarantees.

\subsubsection{Compatibility}
Despite these operational considerations, the proposed framework is fully compatible with existing AP2 deployments. ZTRV does not require changes to the AP2 protocol specification or mandate format. Instead, it can be introduced as an execution-layer enhancement, for example by integrating it into an API gateway, reverse proxy, or service mesh sidecar. This design allows incremental adoption without disrupting existing agent or merchant implementations.

\subsection{Limitations}
This work assumes that the cryptographic primitives and key management infrastructure underlying AP2 mandates remain uncompromised and does not address physical or side-channel attacks. Additionally, while ZTRV enforces strict context binding and consume-once semantics, it cannot prevent authorized misuse scenarios in which a legitimate agent is manipulated such as through prompt injection to obtain and execute a valid mandate for an unintended or malicious purpose. Addressing such semantic or intent-level attacks remains an open challenge beyond the scope of execution-layer verification. These limitations also highlight the need for complementary defenses at the agent reasoning and intent-validation layers.

\section{Conclusion}

As autonomous agents increasingly participate in the digital economy, ensuring the security of agent-driven payment execution becomes a critical systems and security challenge. This work demonstrates that zero-trust runtime verification can effectively mitigate replay and context-binding failures in mandate-based payment protocols such as AP2, without imposing significant performance overhead.

Through systematic analysis and ablation studies, we show that context-aware binding and consume-once enforcement address distinct and complementary failure modes, and that both are necessary to ensure robust runtime security under agentic concurrency and retries. Our evaluation further indicates that the state required for enforcement is bounded by peak concurrency rather than cumulative transaction history, and that verification overhead remains predictable and dominated by cryptographic operations.

By enforcing authorization properties at execution time rather than relying solely on static specification-level guarantees, the proposed framework bridges the gap between protocol assumptions and runtime reality. These findings suggest that runtime enforcement should be considered a foundational component of secure engineering for autonomous systems \cite{anderson2010security}. Looking forward, future work will investigate the integration of Trusted Execution Environments (TEEs) \cite{niu2025you} to further harden the verifier against physical compromise and strengthen end-to-end trust in agentic payment execution.

\section*{Disclaimer}
This work is for research purposes only. The views expressed are those of the authors and do not necessarily reflect the official policy or position of eBay Inc. All experiments were conducted in a simulated environment.

\bibliographystyle{unsrt}  
\bibliography{references}

\end{document}